\documentclass[%
 aip,
 amsmath,amssymb,
 reprint,%
floatfix]{revtex4-1}

\usepackage{graphicx}
\usepackage{dcolumn}
\usepackage{bm}
\usepackage{pgf}
\usepackage{tikz} 

\usepackage[utf8]{inputenc}
\usepackage[T1]{fontenc}
\usepackage{mathptmx}
\usepackage{etoolbox}

\usepackage{hyperref}
\hypersetup{
    colorlinks=true,
    linkcolor=blue,      
    citecolor=blue,      
    urlcolor=blue        
}

\makeatletter
\def\@email#1#2{%
 \endgroup
 \patchcmd{\titleblock@produce}
  {\frontmatter@RRAPformat}
  {\frontmatter@RRAPformat{\produce@RRAP{*#1\href{mailto:#2}{#2}}}\frontmatter@RRAPformat}
  {}{}
}%
\makeatother
\begin{document}

\preprint{AIP/123-QED}


\title[Uhlenbeck-Ford model in 2D]{The Uhlenbeck–Ford model in two dimensions: Reference system for fluid-phase free-energy calculations}

\author{Samuel Cajahuaringa}
\affiliation{Dipartimento di Fisica, Universit\`a di Trieste, I-34151 Trieste, Italy}
\author{Rodolfo Paula Leite}%
\affiliation{Instituto de F\'{i}sica Gleb Wataghin, Universidade Estadual de Campinas, UNICAMP, 13083-859 Campinas, S\~{a}o Paulo, Brazil}%
\author{Maurice de Koning}
\affiliation{Instituto de F\'{i}sica Gleb Wataghin, Universidade Estadual de Campinas, UNICAMP, 13083-859 Campinas, S\~{a}o Paulo, Brazil}
\affiliation{Center for Computing in Engineering and Sciences, Universidade Estadual de Campinas, UNICAMP, 13083-861 Campinas, S\~{a}o Paulo, Brazil}%

\date{\today}

\begin{abstract}
We investigate the Uhlenbeck--Ford (UF) model as a reference system for free-energy calculations in two-dimensional (2D) fluids. The 2D virial coefficients are computed exactly up to tenth order and combined with molecular simulation data to construct highly accurate numerical representations of the equation of state and the excess Helmholtz free energy. We then determine the phase diagram of the model in order to establish the thermodynamic stability limits of the fluid phase and thereby identify the range of applicability of the UF model as a fluid reference system. In the course of this analysis, we identify the solid, hexatic, and fluid phases, and show that the fluid remains the only thermodynamically stable phase, independent of density, for scaling parameters up to $p\lesssim 70$. Finally, we demonstrate the practical applicability of the 2D UF model as a reference system through thermodynamic integration calculations of the free energy of a two-dimensional Lennard--Jones fluid.

\end{abstract}

\maketitle

\section{\label{sec:introduction}Introduction}
Advances in nanofabrication and atomic-resolution characterization have transformed the study of low-dimensional materials, making it possible to isolate and probe two-dimensional (2D) systems with unprecedented detail.~\cite{Alam2021,Algara2015,Buchner2017,AlMutairi2024,Vidya2018,Bui2025} In contrast to three-dimensional (3D) materials, where solid--liquid transitions are generally first order, two-dimensional systems exhibit a much richer spectrum of phase behavior, including defect-mediated melting, intermediate hexatic phases,~\cite{Kosterlitz1973,Halperin1978,Nelson1979,Young1979,Kapil2022} and confinement-induced structures. Such phenomena have been observed experimentally in colloidal and granular monolayers,~\cite{Gasser2010,Han2008,Zaluzhnyy2017} atomically thin materials,~\cite{Bui2025} and in simulations of systems described by a wide range of interaction potentials.~\cite{Prestipino2011,Zu2016,Russo2017,Kapfer2015,Hajibabaei2019,Toledano2021,Li2020,Coto2024} These developments highlight the need for accurate theoretical methods capable of predicting phase stability and phase boundaries in two-dimensional systems.

A central challenge in the theoretical study of two-dimensional phase transitions is the accurate calculation of free energies, which determine the relative stability of competing phases. The thermodynamic integration (TI) approach~\cite{Kirkwood1935,FrenkelSmit} determines free-energy differences by constructing a reversible thermodynamic path between the system of interest and a reference system of known free energy and evaluating the reversible work performed along this path. In principle, the method is exact, provided a suitable reference system is available. The choice of such a reference depends on the phase under consideration and the properties of interest. For solids, the Einstein crystal provides a natural reference because its free energy is known analytically.~\cite{FrenkelSmit,Freitas2016,deKoning1996,Khanna2021} For fluids in three dimensions, the Uhlenbeck--Ford (UF) model has emerged as an attractive reference system.~\cite{Leite2016,Leite2019}
Originally introduced as a theoretical model for the analytical evaluation of virial coefficients,~\cite{Uhlenbeck1962,McQuarrie2000,Leite2016} the UF model is based on an ultrasoft purely repulsive pair potential with a logarithmic divergence at the origin. A distinctive feature of the model is that its thermodynamic properties are completely determined by a single dimensionless density variable, while temperature enters only as a scaling parameter.~\cite{Leite2016,Leite2019} Previous studies have shown that these properties make the UF model an accurate and computationally efficient reference system for fluid-phase free-energy calculations in three dimensions.~\cite{Leite2019,Cajahuaringa2022} This naturally raises the question of whether the UF model can play a similar role in two dimensions. However, its thermodynamic properties and phase behavior in two dimensions have not yet been systematically characterized, and its suitability as a reference system for two-dimensional fluids therefore remains unexplored.

In this work, we address this gap through a comprehensive investigation of the two-dimensional UF model. We determine the exact virial coefficients up to tenth order and combine them with molecular dynamics simulations to construct a highly accurate equation of state (EOS) and excess Helmholtz free-energy expression. Building on this framework, we develop numerical representations of the thermodynamic properties for different values of the scaling parameter $p$. We then determine the phase diagram of the model in order to establish the thermodynamic stability limits of the fluid phase and thereby identify the range of applicability of the UF model as a reference system. This analysis also reveals the occurrence of an intermediate hexatic phase over a limited region of the phase diagram. Finally, we demonstrate the practical usefulness of the UF model as a reference system by applying it to thermodynamic integration calculations of the free energy of a two-dimensional Lennard--Jones fluid.

The remainder of this paper is organized as follows. In Sec.~\ref{section:models}, we describe the computational methodology, including the UF model and the thermodynamic integration approach used to calculate free-energy differences. In Sec.~\ref{section:results}, we present the results, beginning with the EOS and corresponding excess Helmholtz free-energy representations in Sec.~\ref{section:EOS}, followed by the phase diagram defining the stability limits of the fluid phase in Sec.~\ref{section:phase_diagram}, and finally the application of the two-dimensional UF model as a reference system for TI calculations of the free energy of the two-dimensional Lennard--Jones (LJ) fluid in Sec.~\ref{section:application}. We conclude with a summary in Sec.~\ref{section:conclusion}.
\section{\label{section:models}Computational methodology}
\subsection{\label{section:UF}The Uhlenbeck-Ford model }
The UF model is defined by the interatomic pair potential~\cite{Uhlenbeck1962,Leite2016}
\begin{equation}
U_{\mathrm{UF}}(r)=-\frac{p}{\beta}
\ln\left[1-e^{-(r/\sigma)^{2}}\right],
\label{eqn:UFM_pot}
\end{equation}
where $\beta\equiv(k_{B}T)^{-1}$, $\sigma$ is an interaction length scale and $p>0$ is a dimensionless scaling factor. The interaction is smooth, purely repulsive, diverges logarithmically as $r\to 0$, and decays rapidly with increasing $r$. Due to the explicit temperature dependence of the potential, $p$ controls the effective strength of the interaction and can be interpreted as a dimensionless inverse effective temperature $T^{*}\equiv 1/p$.

A distinguishing feature of the UF model is its analytical tractability: its virial coefficients can, in principle, be evaluated exactly for arbitrary spatial dimension $d$~\cite{Baram1991} and any positive integer $p$.~\cite{Leite2016}. In particular, the virial expansion in $d$ dimensions can be written in the form~\cite{Leite2016}
\begin{equation}
    \beta bP=x+\sum_{n=2}^{\infty}\widetilde{B}^{(p)}_{n}x^{n}
    \label{eqn:virial_P}
\end{equation}
with $x$ the dimensionless density $x\equiv b\rho$,~\cite{Leite2016} where $b\equiv\tfrac{1}{2}(\pi\sigma^{2})^{d/2}$ and $\rho\equiv N/A$ is the two-dimensional number density, $A$ being the area occupied by the system. The Mayer cluster expansion~\cite{Mayer1940,McQuarrie2000} then gives the virial coefficients $\widetilde{B}^{(p)}_{n}$ as~\cite{Leite2016}
\begin{align}
    \widetilde{B}^{(p)}_{n}&=\frac{(1-n)2^{n-1}}{n!}\sum_{\mathcal{G}}w(\mathcal{G})\Bigg(\sum_{k_{1}=1}^{p}\dots\sum_{k_{m}=1}^{p}(-1)^{k_{1}+\dots+k_{m}} \nonumber \\
    &\times\binom{p}{k_{1}}\dots\binom{p}{k_{m}}\mbox{det}\left[\textbf{M}(\mathcal{G},\{k_{1},\ldots,k_{m}\})\right]^{-d/2}\Bigg), 
    \label{eqn:Bn}
\end{align}
where $w(\mathcal{G})$ denotes the weight of the unlabeled biconnected graph $\mathcal{G}$ containing $n$ vertices and $m$ edges and $\mbox{det}\left[\mathbf{M}(\mathcal{G},\{k_{1},\ldots,k_{m})\right]$ is the determinant of the minor matrix associated with the weighted graph obtained by assigning weights $k_l$ to the $m$ edges of $\mathcal{G}$.~\cite{Leite2016} In particular, the $\widetilde{B}^{(p)}_{n}$ are independent of temperature, a direct consequence of the fact that the energy scale of the UF potential is proportional to the thermal energy $k_B T$. The corresponding excess Helmholtz free energy per particle is then given by~\cite{Leite2016}
\begin{equation}
    \frac{\beta F^{\mathrm{exc} (p)}_{\mathrm{UF}}(x)}{N} =\sum_{n=1}^{\infty}\frac{\widetilde{B}^{(p)}_{n+1}}{n}x^{n}.
    \label{eqn:Fexec_p}
\end{equation}
\subsection{\label{section:TI}Nonequilibrium thermodynamic integration}

Accurate determination of free energies is essential for characterizing phase stability and thermodynamic properties of condensed-matter systems. Given that free-energies are thermal quantities that cannot be expressed as ensemble averages,~\cite{FrenkelSmit} they are usually obtained as free-energy differences with respect to a reference system for which the free energy is known. In molecular simulations, these differences are often determined using the thermodynamic integration (TI) method which is based on constructing a parametrized Hamiltonian $H(\lambda)$ that continuously interpolates between the Hamiltonian of the system of interest $H_{\text{int}}$ and the system of reference $H_{\text{ref}}$
\begin{equation}
    H(\lambda) = (1 - \lambda) H_{\text{ref}} + \lambda H_{\text{int}},
    \label{eq:H_lambda}
\end{equation}
where $\lambda$ is a coupling parameter varying between 0 and 1. The Helmholtz free-energy difference $\Delta F$ between these two states is equal to the reversible work performed along a quasistatic path:
\begin{equation}
    \Delta F = F_{\text{int}} - F_{\text{ref}} = \int_{0}^{1} \left\langle \frac{\partial H(\lambda)}{\partial \lambda} \right\rangle_{\lambda} d\lambda \equiv W^{0\to1}_{\text{rev}},
    \label{eq:TI_integral}
\end{equation}
where $\langle \dots \rangle_{\lambda}$ denotes the equilibrium ensemble average at a specific value of $\lambda$, and $\partial H / \partial \lambda$ is the thermodynamic driving force. In practice, evaluating Eq.~(\ref{eq:TI_integral}) requires performing a series of independent equilibrium simulations at discrete $\lambda$-windows, followed by numerical integration. 
This approach is computationally rather demanding, however, due to the need for multiple independent equilibrium simulations. To reduce the computational cost, we employ a non-equilibrium approach to the computation of the reversible work,~\cite{deKoning2005,Freitas2016} in which the coupling parameter $\lambda=\lambda(t)$ is explicitly time-dependent during a simulation, varying between 0 and 1 over a finite switching time interval of duration $t_s$. 

In this nonequilibrium thermodynamic integration (NETI) approach the reversible work integral is estimated in terms of the dynamical work $W^{0\to1}_{\text{dyn}}$, calculated by integrating the instantaneous driving force along the non-equilibrium phase-space trajectory $\Gamma(t)$:
\begin{equation}
    W^{0\to1}_{\text{dyn}} = \int_{0}^{t_s} \frac{d\lambda}{dt} \left. \frac{\partial H(\lambda)}{\partial \lambda} \right|_{\Gamma(t)} dt.
    \label{eq:W_dyn}
\end{equation}
Because the finite-time switching process is intrinsically irreversible, dissipative entropy is produced, which gives rise to a systematic error.~\cite{deKoning2005,Freitas2016} However, if the switching process is sufficiently slow such that the system remains in the linear-response regime, this error can be eliminated by combining the results of forward ($W^{0\to1}_{\text{dyn}}$) and backward ($W^{1\to0}_{\text{dyn}}$) switching processes. The unbiased estimator for the free-energy difference is then given by:~\cite{deKoning2005,Freitas2016}
\begin{equation}
    \Delta F = \frac{1}{2} \left[ \overline{W}_{\text{dyn}}^{0\to1} - \overline{W}_{\text{dyn}}^{1\to0} \right],
    \label{eq:unbiased_estimator}
\end{equation}
where the bar indicates an average over independent realizations of the nonequilibrium switching process. 

Using the UF fluid as the reference system in Eq.~(\ref{eq:H_lambda}), the free energy of a fluid of interest can then be determined using the forward and backward dynamical work estimators according to Eq.~(\ref{eq:unbiased_estimator}),
\begin{equation}
    F_{\rm int} = F_{\rm ig} + F_{\text{UF}}^{\text{exc} (p)}+ \frac{1}{2} \left[ \overline{W}_{\text{dyn}}^{0\to1} - \overline{W}_{\text{dyn}}^{1\to0} \right]
\end{equation}
where $F_{\text{UF}}^{\text{exc} (p)}$ is the known UF excess Helmholtz free energy and $F_{\rm ig}$ is the ideal gas contribution.
\section{\label{section:results}Results}

\subsection{\label{section:EOS}Equations of state and excess Helmholtz free energies of the 2D UF model}
Using the virial expansion of Eq.~(\ref{eqn:virial_P}), we first analyze the equation of state by computing the virial coefficients $\widetilde{B}^{(p)}_{n}$ for $p=1$, with the expansion truncated at $n=10$, beyond which the calculation becomes computationally prohibitive due to the factorial growth in the number of graphs.~\cite{Leite2016} To this end, we evaluate Eq.~(\ref{eqn:Bn}) by explicitly generating all unlabeled biconnected graphs $\mathcal{G}$ for a given $n$ using the \texttt{nauty} software package\cite{McKay2014}, and subsequently computing their corresponding weights. The resulting virial coefficients up to $\widetilde{B}_{10}$ are summarized in Table~\ref{tab:virial}.
\begin{table}[b]
\caption{\label{tab:virial}Numerical representation of the exact virial coefficients for the UF model up to $\widetilde{B}^{(p)}_{10}$ for $p=1$.}
\begin{ruledtabular}
\begin{tabular}{cdr}
n & \centering{\text{Graphs}} & $\widetilde{B}_{n}$ \\
\hline
2 & 1 & 1 \\
3 & 1 & 0.44444444444444444444 \\
4 & 3 & -0.06250000000000000000 \\
5 & 10 & -0.12732582972582972583 \\
6 & 56 & 0.03658348873072061913 \\
7 & 468 &   0.07201486005517357001 \\
8 & 7123 & -0.02465550662234846981 \\
9 & 194066 & -0.05042432226902967031\\
10 & 9743542 & 0.01774803730504760993 \\
\end{tabular}
\end{ruledtabular}
\end{table}

\begin{figure}[tbp]
    \centering
    \includegraphics[width=1.0\linewidth]{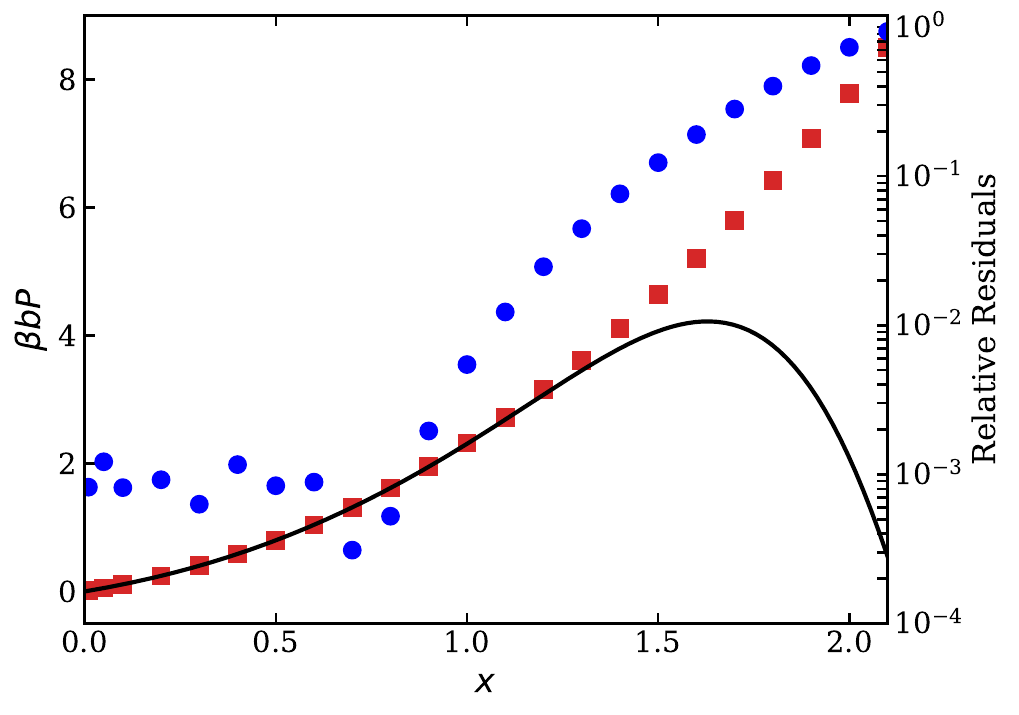}
    \caption{Comparison of the reduced pressure as a function of the dimensionless variable x for the 2D UF model with $p=1$, as obtained from the exact virial expansion truncated after $\widetilde{B}^{1}_{10}$ using the coefficients listed in Table~\ref{tab:virial} (solid line) and from MD simulations (red squares). The relative residuals are shown as blue circles. The statistical uncertainty of the MD averages is smaller than the symbol size over the entire range of $x$.}
    \label{fig:eos_ufm_2D}
\end{figure}
To assess the accuracy of the truncated virial expansion, we compare it to pressure values obtained from MD simulations of the 2D UF model using the \verb|LAMMPS| code~\cite{Plimpton1995,Thompson2022}. The UF parameters were chosen as $\sigma=\sqrt{2/\pi}$, which corresponds to $b=1$, and $\beta=1$, with a cut-off radius of $r_{c}=5\sigma$. Simulations were performed in a square periodic cell containing $10^4$ particles. The equations of motion were integrated using the Langevin thermostat~\cite{Schneider1978} for $10^{6}$ time steps with $\Delta t=0.001$ (reduced units) and a damping time of $100\Delta t$. The mean dimensionless pressure $\beta bP$ was then determined as a function of the dimensionless density variable $x$ from a series of simulations at different number densities. The results, shown in Fig.\ref{fig:eos_ufm_2D}, are in good agreement with the truncated virial expansion for $x\lesssim1$, with relative errors below $10^{-2}$. At larger values of $x$, however, the virial expansion rapidly deviates from the MD data. This behavior is qualitatively similar to that observed in the 3D model~\cite{Leite2016} and reflects the alternating signs of the virial coefficients together with the slow decay of their magnitudes. As noted in Ref.~\onlinecite{Leite2016}, this limitation becomes even more pronounced for integer scaling parameters $p>1$. Consequently, despite its analytical form, the virial expansion is impractical for quantitative applications. 

Following Ref.~\onlinecite{Leite2016}, we construct alternative representations of the EOS for several integer values of $p$ by interpolating the MD data for the reduced pressure $\beta b P$ as a function of $x$ using cubic splines. The corresponding excess Helmholtz free-energy per particle is then determined by direct integration of the thermodynamic relation~\cite{FrenkelSmit} 
\begin{equation}
\frac{\partial}{\partial \rho}\left(\frac{\beta F^{\rm exc}}{N}\right)
=\frac{\beta P}{\rho^2}-\frac{1}{\rho}.
\label{eq:TI}
\end{equation}
which, upon expressing the density in terms of the dimensionless variable $x=b\rho$, gives
\begin{equation}
\frac{\beta F^{\rm exc (p)}_{\rm UF}(x)}{N}
= \int_0^x \frac{1}{x'} \left[ \frac{\beta b P^{\rm (p)}(x')}{x'}-1 \right]dx',
\end{equation}
with $N$ the number of particles in the system. 

The cubic spline representations for the EOS and excess Helmholtz free energies have been determined for $p=1, 25, 50$ and $70$ and implemented in the python script \texttt{2DufGenerator.py} available on the Materials Cloud~\cite{Data_available}.

As an independent determination of the excess Helmholtz free energy, we also employ the NETI approach described in Sec.~\ref{section:TI}. To this end, we introduce the parametric Hamiltonian 
\begin{equation}
{H(\lambda)}=\lambda H_{\rm UF}+(1-\lambda) H_{\rm ig}=K+\lambda U_{\rm UF},
\label{path}
\end{equation}
where $K$ is the kinetic energy and $\lambda$ varies between 0 and 1. A particularly attractive feature of the UF model is that, unlike many other pair potentials, the generalized force associated with the coupling parameter, 
\begin{equation}
\partial H/\partial \lambda = U_{\rm UF}
\end{equation} 
remains finite throughout the integration path, including the ideal-gas limit $(\lambda=0)$. For many commonly used pair potentials, particle overlaps sampled in the ideal gas give rise to divergent contributions to the driving force, rendering direct thermodynamic integration impractical. In contrast, the logarithmic divergence of the UF potential as $r\to 0$ is sufficiently weak for the ensemble average of $U_{\rm UF}$ to remain finite everywhere on this path. In particular, the mean value of the driving force in the 2D ideal gas limit is given by~\cite{Hansen2006}
\begin{eqnarray}
\nonumber
\frac{\left\langle U^{(p)}_{\rm UF}\right\rangle}{N}&=& \pi \rho\int_0^{\infty}r g(r) U_{\rm UF}(r) \,dr \\
\nonumber
&=& -\frac{\pi p \rho}{\beta}\int_0^{\infty} r \ln\left(1-e^{-(r/\sigma)^2}\right) \,dr \\
&=& \frac{\pi^2 \,x \,p}{6\,\beta},
\end{eqnarray}
where we have used $g(r)=1$ as the radial distribution function for the ideal gas.

\begin{figure}[tbp]
    \centering
    \includegraphics[width=1.0\linewidth]{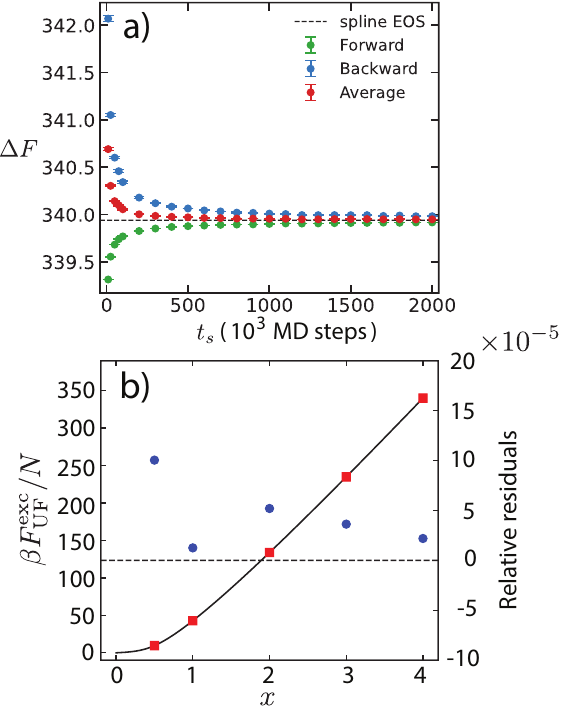}
    \caption{a) Convergence of nonequilibrium work estimator a a function of the switching time $t_{\rm s}$ for $p=70$ and $x=0.4$ b) Comparison of the excess Helmholtz free energy as a function of the dimensionless variable $x$ for the 2D UF model with $p=70$, obtained from the cubic-spline equation of state (solid line) and the NETI results (red squares). The relative residuals are shown as blue circles. The statistical uncertainty of the NETI data is smaller than the symbol size over the entire range of $x$.}
    \label{fig:fexc_ufm_2D}
\end{figure}
Figure~\ref{fig:fexc_ufm_2D} compares the excess Helmholtz free energy obtained from the cubic-spline equation of state with the converged results of the NETI simulations based on Eq.~(\ref{path}) for $p=70$ over a range of densities $x$. The excellent agreement, with relative residuals below $2\times10^{-4}$ throughout the entire density range, independently validates the cubic-spline equation of state. Validations for the remaining values of $p$ are provided in the Supplementary Material (SM).

\subsection{\label{section:phase_diagram}Phase diagram}
In three dimensions, the UF fluid remains the only thermodynamically stable phase for scaling parameters $p\lesssim100$, regardless of the density $x$,~\cite{Leite2017} making the model a suitable reference system for fluid-phase free-energy calculations. It is therefore natural to ask whether the fluid exhibits a similarly broad stability range in two dimensions. To answer this question, we determine the phase diagram of the 2D UF model by performing molecular dynamics simulations over a broad range of densities and scaling parameters $p$. This enables us to establish the thermodynamic stability limits of the fluid phase.

To establish the phase diagram, we analyze the structural evolution of the system as a function of the scaling parameter $p$. In particular, we determine the melting boundary of the crystalline phase. Since intermediate hexatic phases have been reported in a wide variety of two-dimensional model systems,~\cite{Prestipino2011,Zu2016,Russo2017,Kapfer2015,Hajibabaei2019,Toledano2021,Li2020,Coto2024} we also investigate whether the 2D UF model exhibits such a phase. To this end, we compute the static structure factor 
\begin{equation}
S(\mathbf{k})=\frac{1}{N}\left\langle\left|\sum_{i=1}^{N}e^{i\mathbf{k}\cdot\mathbf{r}_i}\right|^2\right\rangle,
\end{equation} 
where \(\langle\cdots\rangle\) denotes an ensemble average, the translational order parameter
\begin{equation}
\Psi_T=\frac{1}{N}\left\langle\left|\sum_{i=1}^{N}e^{i\mathbf{k}_0\cdot\mathbf{r}_i}\right|\right\rangle,
\end{equation}
with \(\mathbf{k}_0\) the characteristic reciprocal-lattice wave vector estimated from the positions of the peaks in the static structure factor,~\cite{Li2019} and the bond-orientational order parameter
\begin{equation}
\Psi_6=\frac{1}{N}\left\langle\left|\sum_{i=1}^{N}\psi_6(\mathbf{r}_i)
\right|\right\rangle.
\end{equation} 
Here,
\begin{equation}
\psi_6(\mathbf{r}_i) = \frac{1}{N_{\rm nn}(i)}
\sum_{j=1}^{N_{\rm nn}(i)}e^{6i\theta_{ij}}
\end{equation}
is the local bond-orientational order parameter, with \(N_{\rm nn}(i)\) the number of nearest neighbors of particle \(i\) determined from a Delaunay triangulation,~\cite{deBerg2008} and \(\theta_{ij}\) is the angle of the bond connecting nearest-neighbor particles \(i\) and \(j\) with respect to a fixed reference axis. In addition, we compute the bond-orientational correlation function
\begin{equation}
g_6(r)=\langle \psi_6^*(0)\psi_6(r)\rangle,
\end{equation}
where the average is taken over all particle pairs separated by a distance \(r\),
which provides a direct measure of the spatial decay of bond-orientational correlations.

In two-dimensional systems, melting may proceed through the two-stage continuous mechanism predicted by the KTHNY theory.~\cite{Kosterlitz1973,Halperin1978,Young1979,Nelson1979} The solid-to-hexatic transition is characterized by the loss of quasi-long-range translational order through the thermal unbinding of dislocation pairs, while quasi-long-range sixfold bond-orientational order is preserved. The subsequent hexatic-to-liquid transition occurs through the unbinding of disclination pairs, which destroys the remaining orientational order and yields an isotropic liquid. Consequently, the combined analysis of \(S(\mathbf{k})\), \(\Psi_T\), \(\Psi_6\), and \(g_6(r)\) provides a sensitive means of distinguishing among the crystalline, hexatic, and fluid phases.

\begin{figure}[tbp]
    \centering
    \includegraphics[width=1\linewidth]{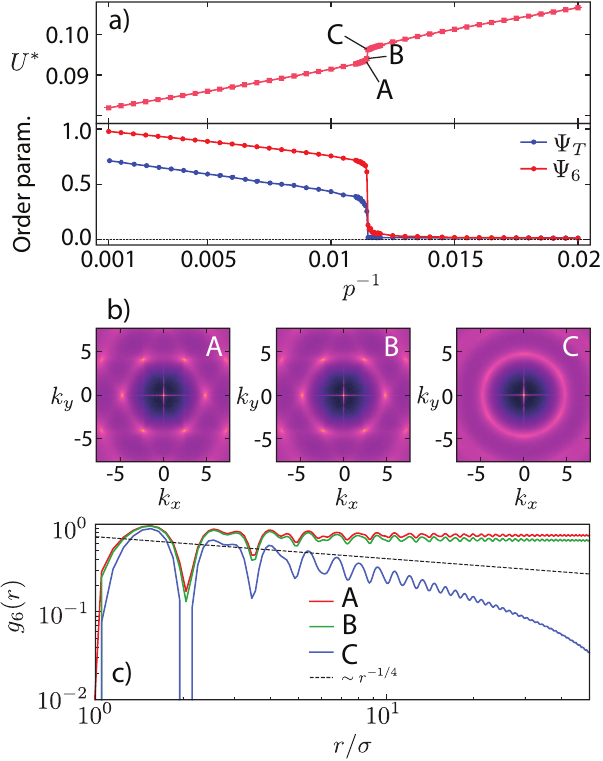}
    \caption{a) Scaled potential energy per particle, \(U^{*}\equiv U_{\rm UF}/(Np)\) (top), and global translational and bond-orientational order parameters, \(\Psi_T\) and \(\Psi_6\) (bottom), as functions of the effective temperature \(T^{*}\equiv 1/p\) for \(x=0.5\). b) Static structure factors across the melting transition at three representative values of \(p\) corresponding to the points labeled A, B, and C in the potential-energy plot in panel a). c) Bond-orientational correlation functions \(g_6(r)\) corresponding to the points labeled A, B, and C in the potential-energy plot in panel a). The dashed line indicates the algebraic decay \(g_6(r)\sim r^{-1/4}\) expected at the hexatic–fluid transition according to KTHNY theory.}
    \label{fig:rho1}
\end{figure}

Using the \verb|LAMMPS| code~\cite{Plimpton1995,Thompson2022} with a Langevin thermostat~\cite{Schneider1978} we performed MD simulations in the canonical ensemble at $T=1$, using a rectangular simulation cell with periodic boundary conditions for a set of densities $\{x_i\}$, with $0.1 \leq x_i \leq 2.0$. The simulations were initialized from a perfect hexagonal crystal containing 11,800 particles at a scaling parameter of $p=1000$, corresponding to an effective temperature $T^{*}\equiv 1/p = 0.001$. The scaling parameter was then gradually reduced to $p=50$, thereby increasing the effective temperature to $T^{*}=0.02$. Initially, $p$ was varied in steps corresponding to $\Delta T^{*}=5\times10^{-3}$, which were subsequently refined to $\Delta T^{*}=5\times10^{-4}$ in the vicinity of the melting transition. At each state point, the system was equilibrated for $5\times10^{5}$ MD steps, followed by a production run of $1\times10^{6}$ steps during which the scaled potential energy per particle  $U^{*}\equiv U_{\rm UF}/(Np)$, the static structure factor $S(\mathbf{k})$, the translational and bond-orientational order parameters, $\Psi_T$ and $\Psi_6$, and the correlation function $g_6(r)$ were computed.

\begin{figure}[t!]
    \centering
    \includegraphics[width=1.0\linewidth]{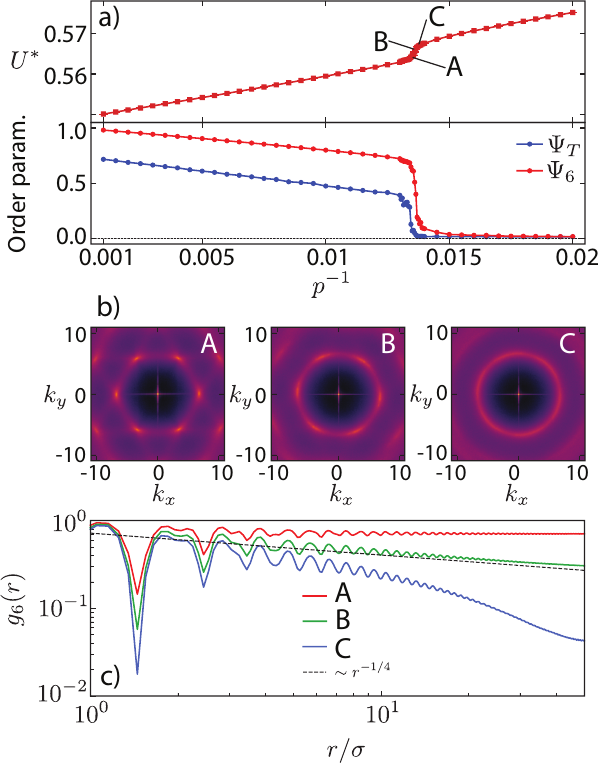}
    \caption{a) Scaled potential energy per particle, \(U^{*}\equiv U_{\rm UF}/(Np)\) (top), and global translational and bond-orientational order parameters, \(\Psi_T\) and \(\Psi_6\) (bottom), as functions of the effective temperature \(T^{*}\equiv 1/p\) for \(x=1.0\). b) Static structure factors across the melting transition at three representative values of \(p\) corresponding to the points labeled A, B, and C in the potential-energy plot in panel a). c) Bond-orientational correlation functions \(g_6(r)\) corresponding to the points labeled A, B, and C in the potential-energy plot in panel a). The dashed line indicates the algebraic decay \(g_6(r)\sim r^{-1/4}\) expected at the hexatic–fluid transition according to KTHNY theory.}
    \label{fig:rho2}
\end{figure}

For \(x<0.1\), the UF model exhibits only the fluid phase, independently of the scaling parameter \(p\) (Fig. S2 in the SM). For \(x\geq0.225\), two distinct melting behaviors are observed. At \(x=0.5\), the scaled potential energy exhibits a clear discontinuity at \(T^{*}\approx0.0114\), indicative of a first-order melting transition, as shown in Fig.~\ref{fig:rho1}(a). Up to this temperature, the static structure factor displays sharp sixfold Bragg peaks characteristic of the crystalline phase. Across the transition, these peaks broaden abruptly while retaining their sixfold symmetry before evolving into the diffuse ring characteristic of the fluid. At the same time, both the translational and bond-orientational order parameters decrease sharply over a narrow temperature interval. The bond-orientational correlation function \(g_6(r)\), shown in Fig.~\ref{fig:rho1}~c), remains finite over the accessible distance range in the crystal but decays rapidly following melting, without displaying behavior consistent with the algebraic decay expected for a hexatic phase.  These observations therefore indicate a direct first-order crystal-to-fluid transition, with no evidence for a stable intermediate hexatic phase.

In contrast, as shown in Fig.~\ref{fig:rho2}, the melting behavior at \(x=1.0\) is qualitatively different. The potential energy varies continuously across the transition region, consistent with continuous melting. The system remains crystalline up to \(T^{*} \approx 0.0134\), followed by an intermediate regime extending approximately from \(T^{*}=0.0136\) to \(T^{*}=0.01375\), during which the sixfold Bragg peaks broaden while remaining clearly discernible. Over the same temperature interval, the translational order parameter \(\Psi_T\) decreases substantially, whereas the bond-orientational order parameter \(\Psi_6\) remains comparatively large, indicating that orientational order persists after translational order has been lost. This interpretation is further supported by the bond-orientational correlation function \(g_6(r)\), whose slow decay in the intermediate regime is consistent with the algebraic behavior expected in a hexatic phase and approaches the KTHNY limiting form \(g_6(r)\sim r^{-1/4}\) near the hexatic–fluid transition. At higher temperatures, the structure factor develops the isotropic diffuse ring of the fluid phase, while \(g_6(r)\) decays more rapidly and no longer displays behavior consistent with quasi-long-range bond-orientational order. Together, these observations provide strong evidence for the two-stage melting scenario predicted by KTHNY theory.~\cite{Kosterlitz1973,Halperin1978,Nelson1979,Young1979}

\begin{figure}[tbp]
    \centering
    \includegraphics[width=1.0\linewidth]{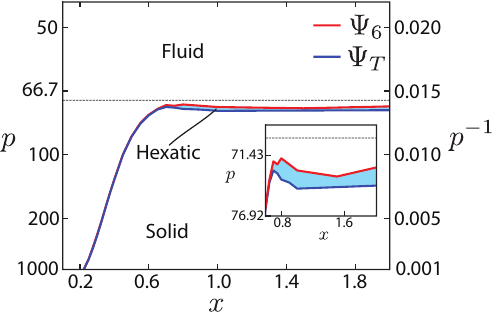}
    \caption{Phase diagram of the two-dimensional UF model. The solid blue and red lines depict the boundaries between the solid and hexatic and between the hexatic and fluid phases, respectively. These were obtained by visually identifying the temperatures at which $\Psi_T$ and $\Psi_6$, respectively, undergo their rapid decrease, corroborated by the evolution of the static structure factor $S(\textbf{k})$. The shaded blue region indicates the hexatic phase. The horizontal black dashed line corresponds to the scaling parameter $p=70$ level. The inset shows an enlarged view of the hexatic phase region.}
    \label{fig:UFM_phase_diagram}
\end{figure}

Extending this analysis to the full density range yields the phase diagram shown in Fig.~\ref{fig:UFM_phase_diagram}. For each density, the solid–hexatic and hexatic–fluid transition temperatures were estimated from the temperature dependence of the translational and bond-orientational order parameters, $\Psi_T$ and $\Psi_6$, respectively, together with inspection of the corresponding static structure factors (see Fig. S2 in the SM). Specifically, the transition temperatures were identified visually from the rapid decrease of the corresponding order parameter, with the structure factor used as a consistency check. The effective melting temperature initially increases with density, reaching a maximum of $T^{*}_{m} \simeq 0.014$, corresponding to $p\simeq 71.43$ at  $x_{m} \simeq 0.8$. At higher densities, the melting temperature remains nearly constant, forming a broad plateau. This behavior contrasts with that of many other soft-core potentials, which exhibit reentrant melting characterized by a decrease in melting temperature with increasing density.~\cite{Prestipino2011,Coto2024,Zu2016} In the 2D UF model, by contrast, the solid phase remains stable over the entire density range investigated, consistent with the behavior previously reported for the three-dimensional model.~\cite{Leite2017}

Evidence for an intermediate hexatic regime is confined to a relatively narrow effective-temperature interval and becomes more pronounced above \(x\simeq0.65\). For \(x<0.65\), the abrupt changes in the scaled potential energy and static structure factor, together with the narrow temperature interval over which the order parameters vary, are consistent with a direct first-order melting transition, with at most a very narrow intermediate hexatic regime. At higher densities, these quantities evolve more gradually and the temperature interval over which translational order is lost while bond-orientational order persists becomes wider, consistent with an increasingly well-defined intermediate hexatic phase. A detailed characterization of the nature of this transition has been the subject of previous studies~\cite{Zu2016,Russo2017,Hajibabaei2019,Toledano2021,Kapfer2015} and is beyond the scope of the present work, whose primary objective is to determine the stability limits of the fluid phase. Overall, these results demonstrate that the fluid phase of the two-dimensional UF model remains thermodynamically stable for arbitrarily high densities provided that \(p\lesssim70\), thereby establishing the range of thermodynamic conditions over which the model can be employed as a fluid reference system.

\subsection{\label{section:application}Application: Helmholtz free energy of the two-dimensional Lennard--Jones fluid}

To demonstrate the practical applicability of the two-dimensional UF model as a reference system, we compute the Helmholtz free energy of the two-dimensional Lennard--Jones (LJ) fluid. Throughout this section, all quantities are reported in reduced LJ units, with distance \(r^{*}=r/\sigma\), number density \(\rho^{*}=\rho\sigma^{2}\), temperature \(T^{*}=k_{\rm B}T/\epsilon\), pressure $P^*=P\sigma^2$ and Helmholtz free energy \(F^{*}=F/\epsilon\). All MD simulations were performed with the \texttt{LAMMPS} code using a periodic simulation cell containing 2500 atoms.~\cite{Plimpton1995,Thompson2022} Temperature control was obtained by the Langevin thermostat.~\cite{Schneider1978} The equations of motion are integrated using a time step of $\Delta t^{*}=0.003$ in reduced units and a damping time scale of 100$\Delta t$. The LJ interactions were modeled using the \texttt{lj/charmmfsw/coul/charmmfsh} pair style~\cite{Pannir2019} with inner and outer switching radii of \(r_{\rm in}^{*}=4.258\) and \(r_{\rm out}^{*}=4.405\), respectively. 

\begin{figure}[tbp]
    \centering
    \includegraphics[width=1.0\linewidth]{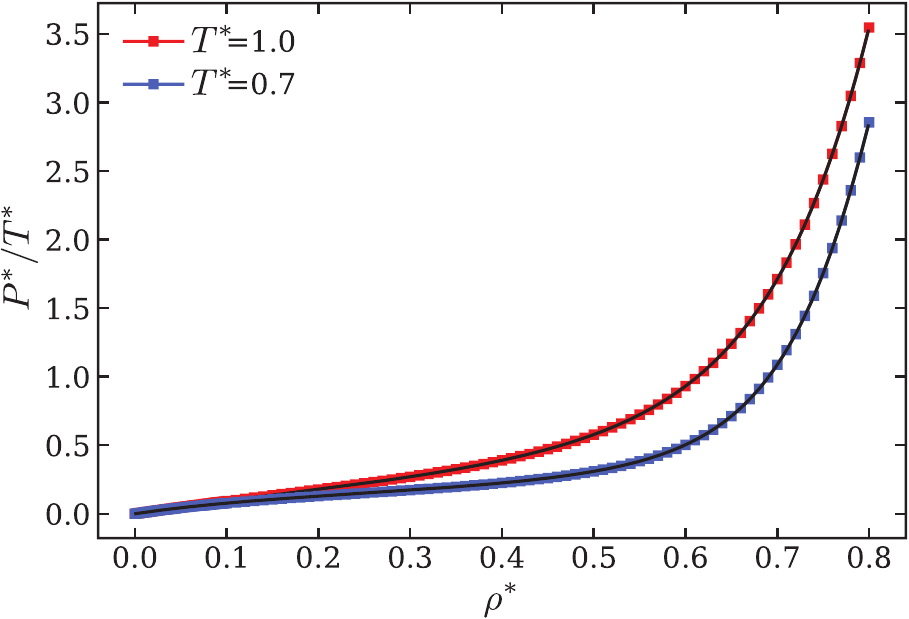}
    \caption{Equations of state of the 2D LJ fluid for $T^*=0.7$ (blue squares) and $T^*=1.0$ (red squares). Black lines represent cubic-spline fits to MD data.}
    \label{fig:EOS_LJ_2D}
\end{figure}

\begin{figure}[bp]
    \centering
    \includegraphics[width=1.0\linewidth]{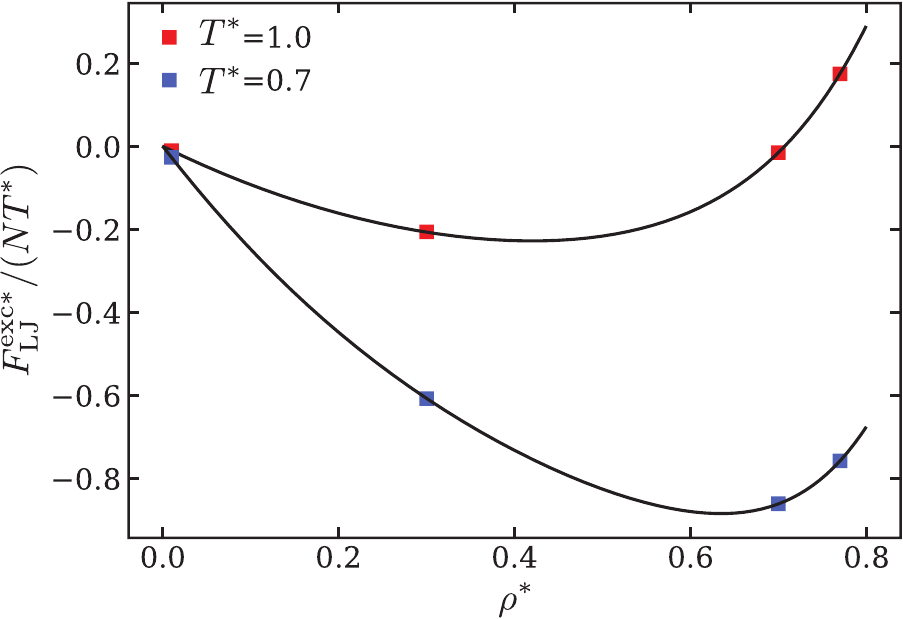}
    \caption{Excess Helmholtz free energy per particle as a function of density for the 2D LJ fluid at \(T^{*}=0.7\) and \(T^{*}=1.0\). The solid lines were obtained by integrating the cubic-spline representations of the equations of state shown in Fig.~\ref{fig:EOS_LJ_2D}. The squares denote results obtained by NETI using the UF model as the reference system.}
    \label{fig:F_exc_LJ_2D_UF}
\end{figure}

To independently validate the Helmholtz free energies obtained from thermodynamic integration, we also determine the equations of state of the 2D LJ fluid at \(T^{*}=0.7\) and \(T^{*}=1.0\). To this end, we perform MD simulations in the NVT ensemble for densities ranging from $0.025$ to $0.8$. Each simulation consists of $10^{7}$ steps of equilibration followed by $10^{8}$  steps of production. This extended production run ensures that the pressure is computed with high accuracy and guarantees decorrelated sampling. Figure~\ref{fig:EOS_LJ_2D} shows the pressure as a function of density obtained from independent MD simulations together with the corresponding cubic-spline representations. These spline fits are subsequently integrated according to Eq.~(\ref{eq:TI}) to obtain the excess Helmholtz free energy, shown by the solid lines in Fig.~\ref{fig:F_exc_LJ_2D_UF}.

\begin{figure}[tbp]
    \centering
    \includegraphics[width=1.0\linewidth]{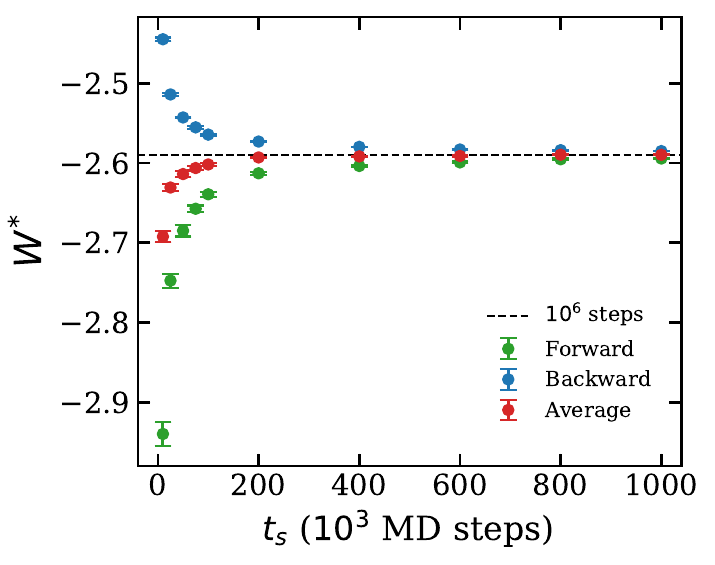}
    \caption{Convergence of irreversible work estimator per particle for NETI process between the LJ and UF models as a function of the number of switching MD steps $t_s$ for $\rho^*=0.3$, $T^*=0.7$ and a scaling factor $p=70$. Before each switching simulation, the system was equilibrated for \(5\times10^{5}\) MD steps. Blue, green and red symbols represent forward, backward and unbiased estimators respectively, obtained using ten independent realizations. Dashed horizontal line represents unbiased result for $t_s=10^6$ MD steps.}
    \label{fig:W_LJ_2D_UF_conv}
\end{figure}

Figure~\ref{fig:W_LJ_2D_UF_conv} shows the convergence of the NETI estimator as a function of the switching time \(t_s\) for \(\rho^{*}=0.3\), \(T^{*}=0.7\), and a scaling factor \(p=70\). The unbiased estimator converges rapidly, differing by less than \(1\times10^{-3}\) from the converged result for \(t_s\gtrsim2\times10^{5}\) MD steps.

Figure~\ref{fig:F_exc_LJ_2D_UF} compares the excess Helmholtz free energies per particle of the 2D LJ fluid obtained from the LJ equations of state shown in Fig.~\ref{fig:EOS_LJ_2D} with those determined by nonequilibrium thermodynamic integration between the LJ and UF models, for densities spanning the fluid phase at \(T^{*}=0.7\) and \(T^{*}=1.0\). The agreement is excellent throughout, with differences of the order of \(10^{-4}\) in all cases (Tab. S1 in the SM), thereby validating the use of the UF model as a reference system for accurate free-energy calculations of two-dimensional fluids.
\begin{table*}[t]
\caption{\label{tab:free_energy_LJ2D} Comparison of the Helmholtz free energy per particle, \(F^{*}\), of the two-dimensional Lennard--Jones fluid obtained from UF-based NETI calculations with values derived from the modified Benedict--Webb--Rubin equation of state (MBWR EOS)~\cite{Reddy1986} and the two-phase thermodynamic (2PT) method.~\cite{Pannir2019} The percentage deviation is calculated relative to the UF-based NETI results. Errors for NETI results correspond to standard errors in the unbiased estimator determined from 10 independent realizations. }
\begin{ruledtabular}
\begin{tabular}{cccccccc}
$\rho^{*}$ & method & $T^{*}=0.45$ & Deviation(\%) & $T^{*}=0.7$ & Deviation(\%) & $T^{*}=1.0$ & Deviation(\%) \\
\hline
0.01 & MBWR EOS & -4.536 & 0.03 & -7.335 & 0.03 & -10.820 & 0.04  \\
 & 2PT(C) & $-4.418\pm0.003$ & 2.63 & $-7.2681\pm0.018$ & 0.88 & $-10.75\pm0.001$ & 0.60 \\
 & 2D UF & $-4.5373\pm0.0002$ & -- & $-7.3326\pm0.0001$ & -- & $-10.8153\pm0.0001$ & --\\
\hline 
0.3 & MBWR EOS & & & -5.386 & 0.51 & -7.646 & 0.47 \\
 & 2PT(C) &  Inhomogeneous state: &  & $-5.115\pm0.003$ & 4.55 & $-7.366\pm0.004$ & 3.21 \\
 & 2D UF &  Liquid/gas coexistence  &  & $-5.3586\pm0.0005$ &  & $-7.6102\pm0.0004$ & -- \\
\hline
0.7 & MBWR EOS & -3.766 & 0.89 & -4.983 & 0.81& -6.622 & 0.77 \\
 & 2PT(C) & $-3.673\pm0.019$ & 1.60 & $-4.939\pm0.014$ & 0.08 & $-6.593\pm0.044$ & 0.32 \\
 & 2D UF & $-3.7328\pm0.0003$ & -- & $-4.9429\pm0.0001$ & -- & $-6.5717\pm0.0001$ & -- \\
\hline
0.77 & MBWR EOS & -3.760 & 0.98 & -4.849 & 0.94 & -6.341 & 0.86\\
 & 2PT(C) & $-3.653\pm0.011$ & 1.89 & $-4.81\pm0.021$ & 0.13 & $-6.31\pm0.003$ & 0.37 \\
 & 2D UF & $-3.72333\pm0.00006$ & -- & $-4.80369\pm0.00009$ & -- & $-6.2867\pm0.0003$ & -- \\
\end{tabular}
\end{ruledtabular}
\end{table*}

Using the NETI approach based on the UF reference system, we compute the Helmholtz free energy of the two-dimensional LJ fluid over a range of densities and temperatures for which the system remains in the homogeneous fluid phase (see Fig. S3 in the SM). The results are then compared with values reported previously for the 2D LJ fluid. In particular, we consider the modified Benedict--Webb--Rubin equation of state (MBWR EOS) proposed by Reddy and O'Shea~\cite{Reddy1986} and the recently developed two-phase thermodynamic (2PT) method.~\cite{Pannir2019} The latter determines the Helmholtz free energy from MD simulations by first obtaining the vibrational density of states from the velocity autocorrelation function and subsequently decomposing it into gas-like (diffusive) and solid-like (vibrational) contributions, from which the entropy and Helmholtz free energy are evaluated. The results are shown in Table~\ref{tab:free_energy_LJ2D}.

The results in Table~\ref{tab:free_energy_LJ2D} demonstrate excellent agreement between the UF-based NETI calculations and the MBWR EOS, with deviations below \(1\%\) for all homogeneous fluid state points considered. At low density (\(\rho^{*}=0.01\)), the MBWR EOS reproduces the UF-based NETI results to within \(0.05\%\), reflecting its excellent accuracy in the dilute regime. At higher densities, the deviations increase, approaching \(1\%\), where the analytical equation of state necessarily becomes less accurate owing to the finite parametrization used in its construction. The comparison with the 2PT method shows a similar level of agreement, although the deviations are generally larger and do not display a systematic dependence on density and temperature. This behavior is consistent with the approximate nature of the 2PT approach, which determines the entropy by decomposing the vibrational density of states into gas-like and solid-like contributions. Such a decomposition is not unique and relies on model assumptions regarding the dynamics of the fluid. In contrast, the NETI method determines the Helmholtz free-energy difference directly through nonequilibrium thermodynamic integration, without introducing model-dependent approximations beyond those inherent to the underlying molecular dynamics description. Consequently, the remaining uncertainties in the NETI results arise only from statistical sampling and finite switching-time effects, both of which can be systematically reduced by increasing the number of realizations and the switching time.

\section{\label{section:conclusion}Conclusions}

In this work, we investigate the thermodynamic properties and phase behavior of the two-dimensional Uhlenbeck--Ford model and show that it provides an efficient and accurate reference fluid for free-energy calculations of two-dimensional fluids. Exact virial coefficients up to tenth order are determined to assess the convergence of the virial expansion. A numerical equation of state is constructed from extensive molecular dynamics simulations, providing an accurate description of the fluid thermodynamics over a broad range of densities. Using structural order parameters, we map out the phase behavior of the 2D UF model and identify an intermediate hexatic phase separating the solid and fluid phases over a broad range of densities. The melting temperature approaches a plateau at high densities, consistent with the behavior previously reported for the three-dimensional UF model.

The resulting phase diagram indicates that the fluid phase remains thermodynamically stable up to approximately $p\lesssim70$ over the entire density range investigated, making the 2D UF model particularly well suited as a reference system for thermodynamic integration calculations of two-dimensional fluids.

As an illustration of its applicability, we employ the 2D UF model as a reference system to determine the Helmholtz free energy of the two-dimensional Lennard--Jones fluid using nonequilibrium thermodynamic integration. The resulting free energies are in excellent agreement with previously reported values throughout the homogeneous fluid region, demonstrating that the 2D UF model provides an accurate and efficient reference system for free-energy calculations in two-dimensional fluids. The present work therefore establishes the 2D UF model as a practical and accurate reference fluid for free-energy calculations of two-dimensional fluids.

\begin{acknowledgments}
S. C. acknowledges the support from the ICSC—Centro Nazionale di Ricerca in HPC, Big Data and Quantum Computing, funded by the European Union—NextGenerationEU (CUP Grant. No. J93C22000540006, PNRR Investimento M4.C2.1.4). Physics Department of the University of Trieste. It was funded through the “Department of Excellence” grant (2023-2027), for providing high-performance computing resources. Part of the calculations was performed at CCJDR-IFGW-UNICAMP. M. K. acknowledges support from the Brazilian agencies CNPq, Capes, Fapesp grant nos. 2016/23891-6, 2018/16572-7, 2019/17874-0, 2020/06896-0 and 2021/03224-3 and the Center for Computing in Engineering \& Sciences - Fapesp/Cepid grant no. 2013/08293-7. Part of the calculations were carried out using resources of the "Centro Nacional de Processamento de Alto Desempenho em São Paulo (CENAPAD-SP)."
\end{acknowledgments}
\section*{Author Declarations}
\subsection*{Conflict of Interest}
The authors have no conflicts to disclose.  
\subsection*{Author Contributions}
\textbf{Samuel Cajahuaringa}: Conceptualization (equal); Data curation (equal); Formal Analysis (equal); Investigation (equal); Methodology (equal); Resources (equal); Software (equal); Validation(equal); Visualization (equal); Writing – original draft (equal); Writing – review \& editing (equal).
\textbf{Rodolfo Paula Leite}: Conceptualization (equal); Data curation (equal); Formal Analysis (equal); Investigation (equal); Methodology (equal); Software (equal); Validation(equal); Visualization (equal); Writing – original draft (equal); Writing – review \& editing (equal)..
\textbf{Maurice de Koning}:  Conceptualization (equal); Supervision (lead); Data curation (equal); Formal Analysis (equal); Investigation (equal); Methodology (equal); Resources (equal); Software (equal);  Validation(equal); Visualization (equal); Writing – original draft (lead); Writing – review \& editing (lead).

\section*{Data Availability Statement}
The data and scripts that support the findings of this study are openly available on the Materials Cloud~\cite{Data_available}.

%

\bibliography{biblio}

\end{document}